\documentclass[12pt,a4paper]{article}
\usepackage{latexsym}
\usepackage{amsmath}
\usepackage{times}

\usepackage[english]{babel}
\usepackage[T1]{fontenc}
\usepackage[ansinew]{inputenc}
\usepackage[pdftex]{graphicx}
\usepackage{subfigure}
\usepackage{hyperref}

\usepackage[svgnames,table,hyperref]{xcolor}
\newcommand{\BJFC}{\bgroup \color{blue}}
\newcommand{\EJFC}{\egroup}

\usepackage{fancyhdr}
\begin{document}
\begin{center}
{\LARGE \bf A Sliding Window Technique for Interactive High-Performance Computing Scenarios} \vspace{0.7cm}

{\large \em Ralf-Peter Mundani$^1$, J\'{e}r\^{o}me Frisch$^2$, Vasco Varduhn$^3$, and Ernst Rank$^1$} \vspace{0.1cm}

{\em \{mundani, frisch, varduhn, ernst.rank\} @ tum.de} \vspace{0.1cm}

{\em $^1$ Technische Universit\"at M\"unchen, Arcisstrasse 21, 80290 M\"unchen, Germany}\\
{\em $^2$ RWTH Aachen University, Mathieustrasse 30, 52074 Aachen, Germany}\\
{\em $^3$ University of Minnesota, 100 Church Street S.E., 55455 Minneapolis, MN, USA}

\end{center}

\noindent \vspace*{10.0mm}\\ {\large \bf Abstract} \\ \newline
Interactive high-performance computing is doubtlessly beneficial for many computational science and engineering applications whenever simulation results should be visually processed in real time, i.\,e.\ during the computation process. Nevertheless, interactive HPC entails a lot of new challenges that have to be solved -- one of them addressing the fast and efficient data transfer between a simulation back end and visualisation front end, as several gigabytes of data per second are nothing unusual for a simulation running on some (hundred) thousand cores. Here, a new approach based on a sliding window technique is introduced that copes with any bandwidth limitations and allows users to study both large and small scale effects of the simulation results in an interactive fashion.

\vspace{0.2in}
\noindent{\bf Keywords:} interactive HPC, sliding window, computational fluid dynamics

\section{Introduction and Motivation}
\label{sec:intro}

Due to recent advances in supercomputing, more and more application domains such as medicine or geosciences profit from high-performance computing (HPC) and come up with (new) complex and computationally intensive problems to be solved. While solving larger problems is just one objective, another one is to solve problems much faster -- even in (or faster than) real time, thus bringing them into the range of interactive computing. Meanwhile many supercomputing centres provide users not only batch access to their HPC resources, but also allow an interactive processing which is in many ways advantageous. The probably most prominent representative of such an approach is \emph{computational steering} \cite{mulder:98} where a simulation back end running on a supercomputer or compute cluster is coupled to a visualisation front end for interaction. Hence, users have the possibility to manipulate certain parameters (geometry, boundary conditions, algorithm control etc.) in order to experience immediate feedback from the running simulation, which \emph{"enhances productivity by greatly reducing the time between changes to model parameters and the viewing of the results \cite{Marshall:CG:90}."} Computational steering has been stated as valuable scientific discovery in the 1987 report of the US National Science Foundation \cite{McCormick:CG:87} and has been researched by various groups worldwide ever since.

At the moment, users can choose from a widespread toolkit of libraries, frameworks, and problem solving environments related to computational steering such as \emph{CUMULVS} \cite{cumulvs}, \emph{RealityGrid} \cite{realitygrid}, \emph{Magellan} \cite{magellan}, \emph{SCIRun} \cite{scirun}, \emph{COVISE} \cite{covise}, or
\emph{Steereo} \cite{jenz:10} -- just to name a few. While those steering approaches differ in the way they provide interactive access to those parameters to be steered/visualised, using check- and breakpoints, satellites connected to a data manager, or data flow concepts, e.\,g., they are usually of limited scope concerning different application domains and/or entail severe code changes. A somewhat generic and `minimal invasive' concept based on the idea of signals and interrupts has been developed by our group that was successfully tested with different applications from various domains (amongst others physics, medicine, and biomechanics) \cite{knezevic:11,knezevic:12} and
that was further integrated into SCIRun \cite{knezevic:12:2}. This concept should now be extended by the idea of a sliding window technique for fast and efficient data transfer between back and front end, especially in case of high-resolution multi-scale simulations.

It should be mentioned at this point, that similar ideas of \emph{sliding windows} can be observed in other disciplines or have been related to other problems. One example would be the so-called one-sided communication proposed in the MPI standard \cite{MPI-2.2}, where a shared memory window can be created and made accessible for remote processes. Even if similar terms or ideas exist in common literature, throughout this paper `sliding window approach' describes a specifically designed data structure running on a high-performance computing system that is connected with a front end machine not necessarily belonging to the same subnet system, in order to visualise parts of the computational multi-resolution domain during runtime.

One major problem within computational steering or interactive computing is the handling of the data transfer between the simulation back end and the interaction front end for visual display of the simulation results. Especially in HPC applications, where a vast amount of data is computed every second, such data advent easily exceeds the capacities, i.\,e.\ bandwidth, of any underlying interconnect\footnote{An apposite description for this matter says \emph{"arithmetic is cheap, latency is physics, and bandwidth is money \cite{hoemmen:10}."}} and, thus, hinders any interactive processing. Hence, sophisticated techniques are inevitable for selecting and filtering the respective data already on the back end in order to cope with physical limitations and to transfer those data only that are really to be visualised. While such an approach is doubtlessly necessary to leverage interactive HPC, on the other hand any selection and/or filtering process discards delicate details of the results that in most cases were computed on a much finer resolution than they are displayed for visualisation. In order to allow users to select any region of interest within the computational domain to be displayed in any granularity (i.\,e.\ the data density of results contributing for visual output) we have developed a sliding window technique that allows for the interactive visual display ranging from coarse (quantitative representation of the entire domain) to fine scales (any details computed on the finest resolution) and which is one important step towards interactive high-performance computing.

By selecting the region of interest -- the window -- on the front end for data transmissions from the back end, an appropriate data structure on the back end becomes necessary, that provides simple and fast access to any subset of the data. Therefore, we have developed a distributed hierarchical data structure that intrinsically supports efficient data access concerning both random choice of subsets and random choice of data density. In order to put everything into practice, we have further implemented a 3D flow solver that carries out its computations on the aforementioned data structure, running in parallel on medium and large size compute clusters and supercomputers, letting us investigate different flow scenarios from simple setups to complex multi-scale problems. In combination with the sliding window data transfer, users have now the choice to retrieve results from a running HPC application to study either large-scale effects (such as flow around a large building or entire city) or small-scale details (such as vortices or local effects) in an interactive fashion. This not only opens the door to many possible scenarios from computational science and engineering, but it also paves the way for interactive HPC which is certainly beneficial for many kinds of optimisation or design problems from various application domains.

This paper is based upon Mundani et al. \cite{Mundani:ParEng:2013}, but the current paper includes the following additional research: usage of the proposed sliding window technique beside solely data selection and transmission also for steering the computation of a large-scale simulation in order to study effects on different scales. The remainder of this paper is as follows. In chapter 2 we discuss the hierarchical data structure and its application for a parallel flow solver, in chapter 3 we introduce the concept of the sliding window and its usage for the interactive data access from a running application. Chapter 4 highlights first results obtained with the sliding window technique including the steering of a large-scale simulation and chapter 5, finally, provides a short conclusion and outlook.

\section{Fundamentals}
\label{sec:impl}

\subsection{Computational Fluid Dynamics}

As implementation example for demonstrating the sliding window data transfer concept we have chosen a parallel computational fluid dynamics (CFD) code currently under development at the Chair for Computation in Engineering at Technische Universit\"at M\"unchen.

The fluid flow computation is governed by the 3-dimensional Navier-Stokes equations for an incompressible Newtonian fluid with no acting external forces:
\begin{eqnarray}
\label{nse_mass_conservation}
\vec \nabla \cdot \vec u & = & 0 \quad ,\\
\label{nse_momentum_conservation}
\frac{\partial}{\partial t} \vec u + \left(\vec u \cdot \vec \nabla \right) \vec u & = & -\frac{1}{\rho} \vec \nabla p + \nu \Delta \vec u  \quad .
\end{eqnarray}
As spatial discretisation, a finite volume scheme is used. The temporal discretisation is realised by a finite difference scheme, namely an explicit Euler method of 2nd order.
A collocated cell arrangement storing all values at the cell centre was chosen together with a pressure oscillation stabilisation method. Numerically, a fractional step method is applied, which is based on an iterative procedure between velocity and pressure during one time step. After an intermediate velocity $v^*$ is computed by neglecting any influence of the pressure field, a Poisson equation is solved, guaranteeing a divergence free pressure distribution. The pressure correction is then added to the intermediate velocity resulting in the velocity at the next time step $v^{n+1}$. Detailed information about the numerical solver as well as validation results can be found in \cite{frisch_csc_2011}.

\subsection{Data structure}

\begin{figure}[htb]
\centering
	\includegraphics[width=11cm]{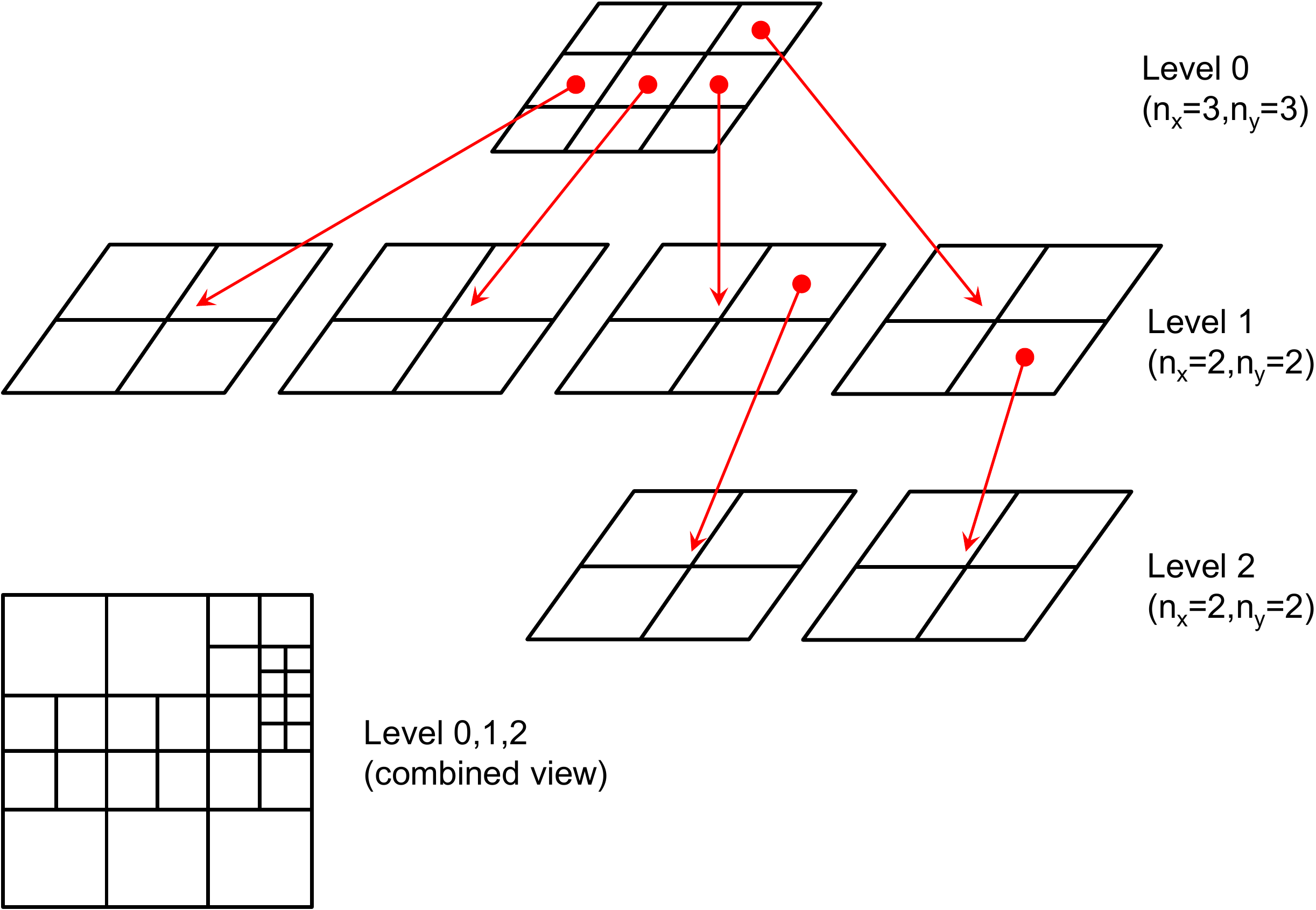}
	\caption{Hierarchical grid construction over different levels using different discretisations (upper part), and a combined view of the `assembled' domain including all grids of this structure (lower left part).}
	\label{fig:fig_grid_levels}
\end{figure}

The data structure of the fluid solver was built from adaptive non-overlapping block-structured equidistant Cartesian grids with a ghost layer halo that favours the sliding window data transfer concept enormously. The code is designed for a full 3D simulation, but for simplifying matters, only 2D grids and domains are drawn in this paper. Figure~\ref{fig:fig_grid_levels} depicts different grid levels with different discretisations and their linkage (upper part) as well as a view of the `assembled' domain consisting of all those grids (lower part). A grid can be divided into $n_x\times n_y\times n_z$ sub-grids with different values for $n_x$, $n_y$, and $n_z$. Furthermore, for level 0 a different discretisation than for levels $\geq 1$ can be chosen. This enables an easy creation of a tunnel shaped domain, e.\,g., while not losing too much `non-computing' cells. A value $n_i=1$ determines that in $i$-direction no refinement takes place over the different hierarchy levels. Obviously, this only makes sense if a `pseudo'-2D computation should be performed using the 3D code. A value of $n_i=2~(i = x,y,z)$ on all levels would describe a regular octree data structure.

\begin{figure}[htb]
\centering
	\includegraphics[width=6cm]{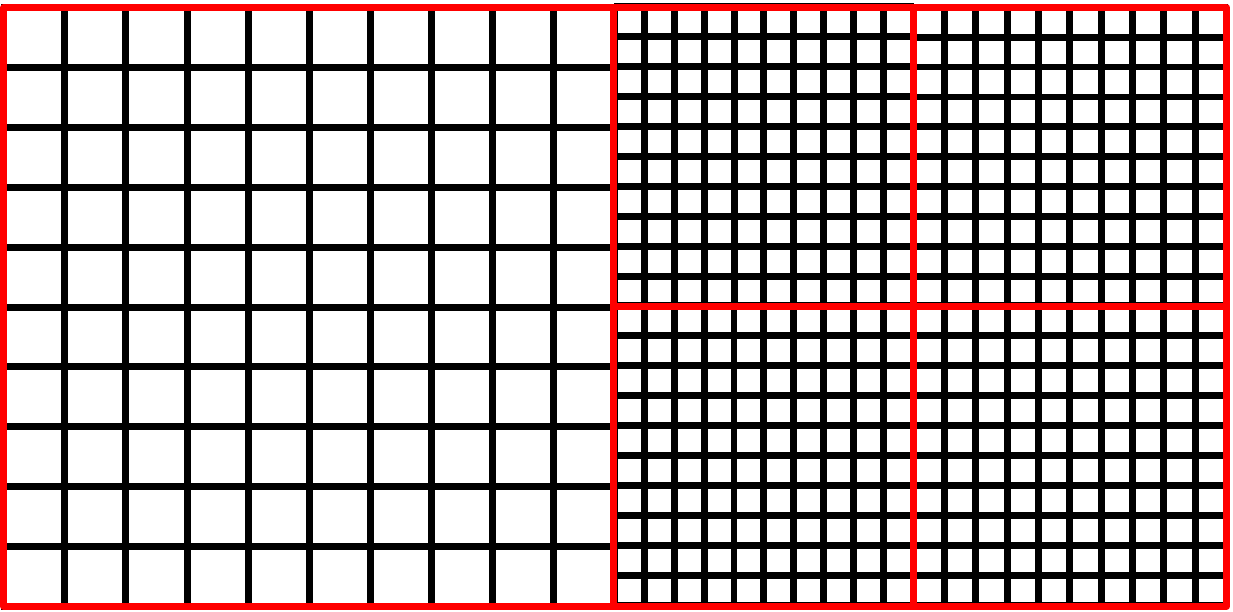}
	\caption{Two adjacent grids (red boundaries) on different levels with their 10$\times$10 cell discretisation (black).}
	\label{fig:fig_grids_with_cells}
\end{figure}

Each grid consists of $nc_x\times nc_y\times nc_z$ cells stored in arrays at grid level that contain the actual computing values in a collocated arrangement (i.\,e.\ in the cell centre). In order to avoid non matching blocks, the following restraint is enforced: $nc_i$ has to be dividable without remainder by $n_i, ~i = x,y,z$ on any level. This ensures that a matching on both sides can be found. In the case of Figure~\ref{fig:fig_grids_with_cells}, the level discretisation is set to $n_x=n_y=2$ with a cell amount of $nc_x = nc_y = 10$. This means, that one cell in the grid on a higher level has two neighbours in the grid on a lower level.

In our hierarchical data structure, grids and corresponding cells are not deleted when a parent grid is refined into child grids. This means, that the cell values are still present, but are not actively involved in the computation step. They merely get filled by interpolated values of the sub-grids during the communication and exchange step briefly described in the following sub-section.

\subsection{Parallel Implementation and Communication Scheme}

\begin{figure}[htb]
\centering
	\includegraphics[width=13cm]{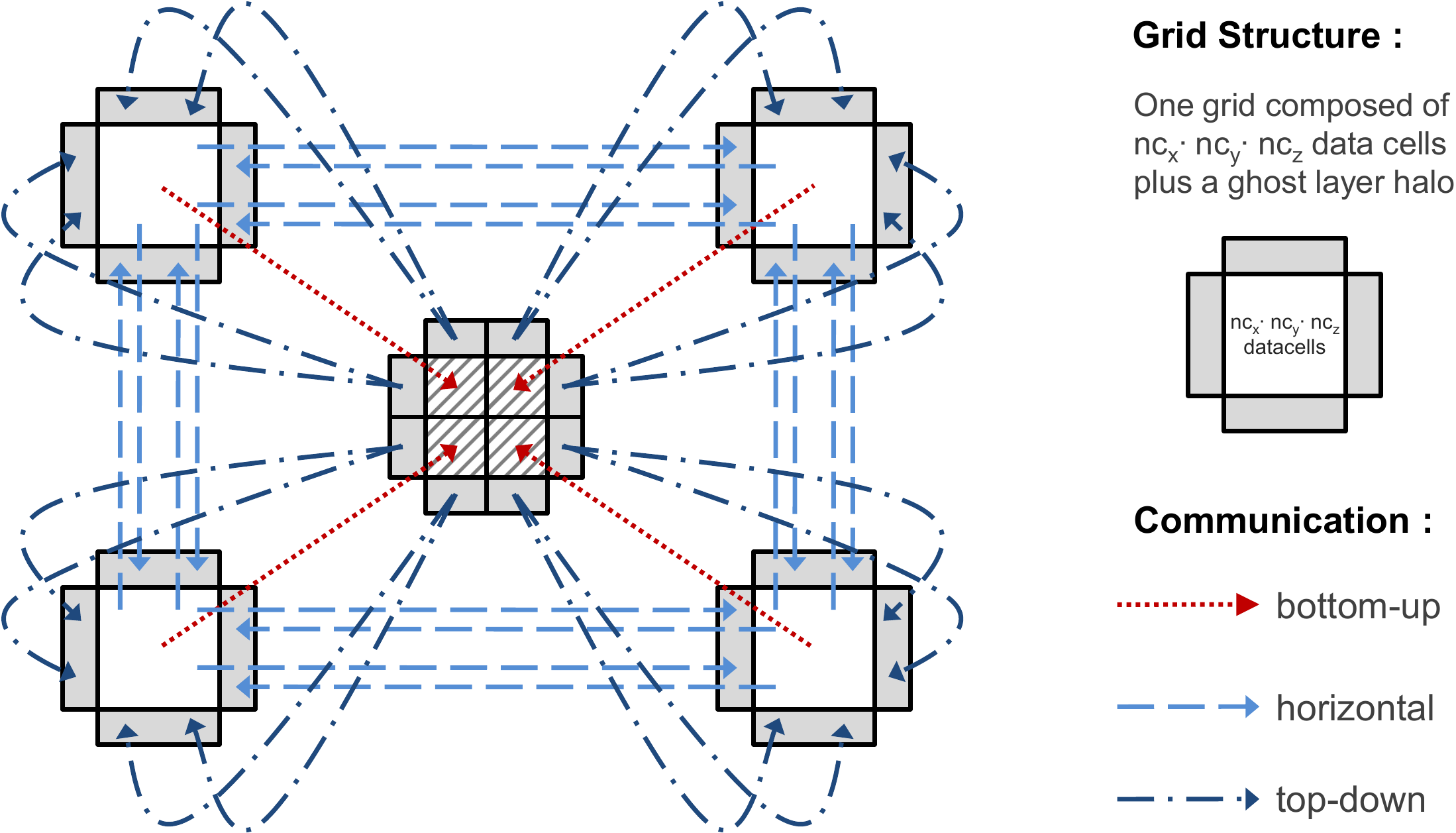}
	\caption{Applied communication scheme for one grid at level 0 (middle grid) and 4 sub-grids on level 1 (outer grids). Each grid is composed of $nc_x\times nc_y\times nc_z$ data cells. The three different communication steps are depicted by different arrows and colours.}
	\label{fig:fig_comm_scheme}
\end{figure}

The parallel implementation is realised by applying a message passing paradigm, namely using MPI \cite{MPI-2.2}. The code is constructed in such a way that a strict separation of computation and communication is ensured, thus enabling programmers with different specialities to focus on their respective area of expertise.

The communication depicted in Figure~\ref{fig:fig_comm_scheme} can be divided into three different steps which require synchronisation in order to ensure a correct data flow. The first communication step can be described as a bottom-up approach. All grids average their data stored in the arrays and send them to the super-grid in order to get stored in the corresponding super-grid cell. Thus, the data travels from the bottom-most grids -- where the actual computation is done -- to the top-most grid which stores the most coarse representation of simulation data. Once the averaged data is stored at the right position, the next step can be applied.

The second step can be categorised as horizontal communication, as here only neighbouring grids on the same level communicate. The neighbouring information can be retrieved from a dedicated neighbourhood server acting as topological repository and answering queries about possible neighbourhood relations in the grid structure \cite{frisch_ispdc_2012}. If a grid has a neighbour, the data is send immediately from this grid to the respective ghost layer halo in the remote grid. Possible problems which will arise due to bottlenecks while increasing the amount of processes are also discussed and remedied in \cite{frisch_ispdc_2012}.

If there was no neighbour present for a certain grid, the ghost layer halo will be filled during the third and final step: the top-down approach. Here data from the parent grid is sent to the ghost layer halo of the sub-grid in order to fill all sides which have not received any data during the horizontal communication step. This procedure also ensures that the data is correctly arriving at grid boundaries, where more than one level difference is present.

The sending and receiving of data is performed by a mixture of blocking and non-blocking calls ensuring a correct synchronisation behaviour. Further detailed information can also be taken from \cite{frisch_synasc_2011}.

\begin{figure}[htb]
\centering
	\includegraphics[width=11cm]{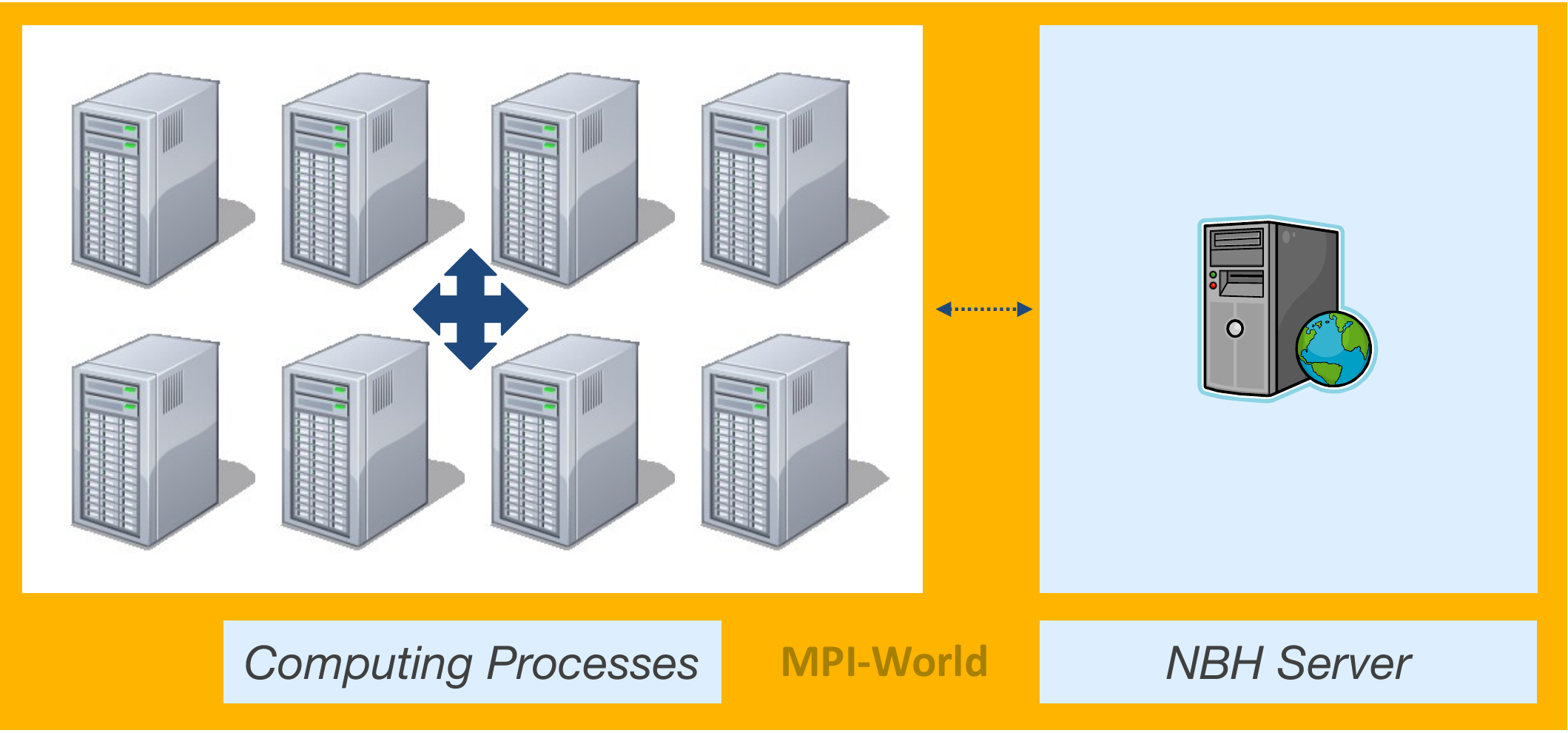}
	\caption{Grouping of processes: Most processes will fulfill computational tasks only, but some might only work on scheduling and organisational tasks. Note that the thickness of the arrows is an indication for the amount of communication between the computing processes themselves and between them and the neighbourhood (NBH) server.}
	\label{fig:fig_mpi_world}
\end{figure}

The different grids will now be distributed among the available processes in the MPI world (Figure~\ref{fig:fig_mpi_world}). For the moment one process is reserved acting as neighbourhood server governing all the relations between the different grids. Please note that the neighbourhood server itself does not contain any computational data, it has a strictly topological view of all grids. 

For distributing the grids, the Morton- or Z-order space filling curve is used, as it can be computed quite fast and easily by bit interleaving \cite{Samet:1984:QRH:356924.356930}. The grids are ordered according to the curve and distributed onto the available computing processes in order to ensure a good initial load distribution. If a lot of adaptive changes are performed in the computational domain, a balancing of the load has to be done at some point to ensure a good load distribution over time.

It can be concluded so far, that the structure and design of our hierarchical grid is very well suited for the sliding window concept as every intermediate level (with interpolated results) is fully present and, thus, a random selection process of data and granularity is strongly supported. A global server can further answer immediately which grids are involved on which processes solely by their IDs only.

\section{Sliding Window}
\label{sec:sliding_window}

\subsection{Concept}
\label{ssec:sliding_window_concept}

The basic idea of the sliding window concept is to limit the necessary data transfer between back and front end by selecting only subsets of the available computed results. Hence, any user should have the possibility to choose both a window, i.\,e.\ a desired region of the computational domain and a desired data density for the respective data to be visualised on the front end. This window can be slid over the computational domain as well as being increased or decreased in size, always determining which subset of data should be selected for the transfer. Whereas the window itself acts as a bounding box for the selection, the data density defines the amount of data points inside that region to be considered for the visual display. Key feature of the entire approach is to keep the data density constant while the window is being moved or resized, thus at any time the same amount of data is transmitted from the back to the front end. This allows us to optimally exploit the underlying network without exceeding any bandwidth limitations or causing unnecessary long transmission delays that would harm the experience of an authentic interactive computing.

In case of high-resolution data, a full window covering the entire computational domain would force the back end (depending on the chosen data density) to skip lots of data points for the data transfer -- for instance selecting only ever fifth or tenth data point in every dimension. Hence, the user would get a qualitative representation without too many details which more or less allows to catch global effects of the running simulation. When resizing the window, i.\,e.\ making it smaller, a smaller region of interest is covered and, thus, in order to keep the data density constant, less data points are to be skipped, providing more and more details for the visual display. The window size can be further decreased until the resolution of the computational domain is met, i.\,e.\ now each single data point inside the selected region is shown on the visual front end. Obviously, the critical questions are how to select the right data points (cf.\ data density) and in case of a parallel simulation how to select the right processes (cf.\ window).

We will see in the next section that our data structure is advantageous for the sliding window concept, as due to the hierarchy of grids together with their update scheme an implementation of this concept is straightforward and questions concerning the right selection of data points and processes are very easy to answer.

\subsection{Implementation}
\label{ssec:sliding_window_implementation}

The implementation of the sliding window concept is based on two major components, namely a server-side collector, responsible for collecting and gathering the requested information on the back end, and a client-side visualisation and interpreter tool at the front end for sending data requests and receiving the corresponding data from the back end.

\subsubsection{Server-Side Collector}

In order to keep the influence between the data collection process and the actual computation processes to a minimum, a special dedicated collector process was introduced into the above described CFD code. The main structure of the collector is an infinite loop listening on a TCP socket on a dedicated port for a single character describing the next task for execution. One possible command from the client-side to the server-side would be to do a sliding window visualisation.

\begin{figure}[htb]
\centering
	\includegraphics[width=\textwidth]{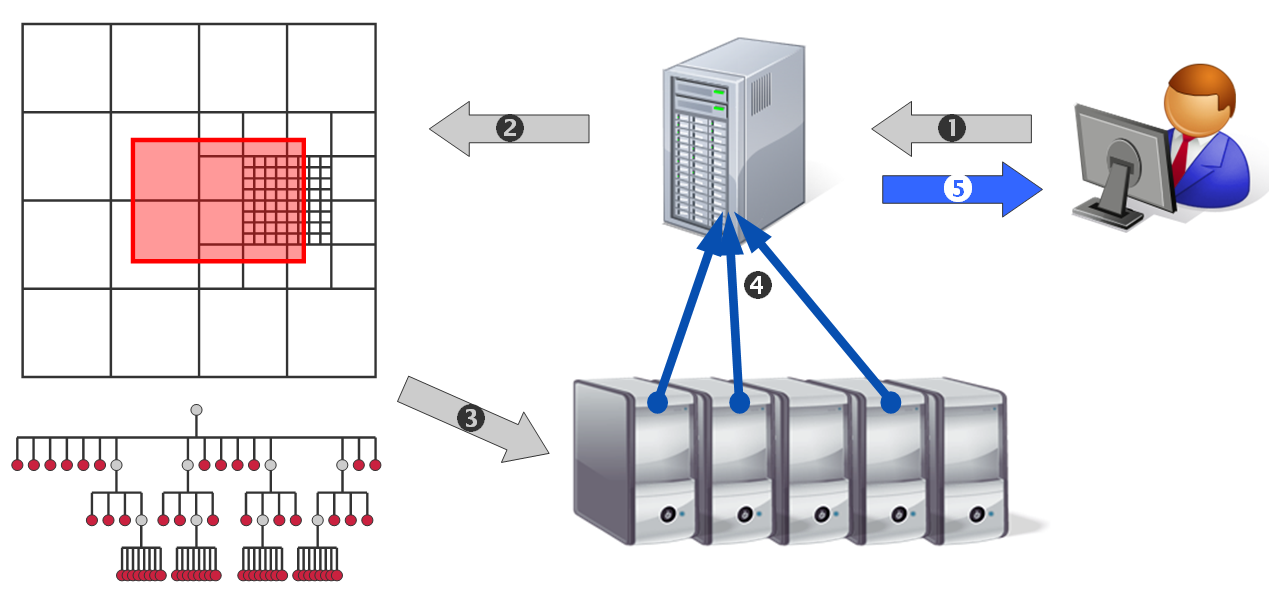}
	\caption{Different steps of a sliding window query: first, the client issues a request to the collector (1) that forwards it to the neighbourhood server for determining all involved grids (2), then the neighbourhood server informs the respective computation processes (3) to send the required data to the collector (4) which returns a formatted stream to the client (5).}
	\label{fig:sliding}
\end{figure}

Once the client has sent the task initialisation, it will continue sending the necessary information, such as visualisation data type, maximum amount of cells and the visualisation bounding box to the server. The collector receives this information and submits a query to the neighbourhood server, which has a topological and geometrical overview of the stored grids. The neighbourhood server performs for every grid an intersection test along the $x$, $y$, and $z$ axis of the grid's bounding box and the transmitted sliding window bounding box. If the grid touches or lies completely inside the sliding window box, it is marked and added to the treatment list. If it is completely outside, the grid and all its sub-grids are ignored henceforth. The order of traversal is given by the same space filling curve used for load distribution. Thus, the grid identification is executed from the coarsest top grids to the more and more fine grids in deeper levels.

As soon as the maximum amount of cells to be displayed is reached and the relevant grids have been identified, the neighbourhood server orders the corresponding processes containing the grids to send a data stream to the collector process. For this, MPI messages using the very fast dedicated cluster interconnect are exchanged. The collector then reforms and compresses the data in a way that it can be sent over a possibly slower TCP connection to the client. During the time the send operation is executed, the computing processes can continue with their regular simulation task without being blocked by the server-to-client transmission. A full pass of such a sliding window query is presented in Figure~\ref{fig:sliding}.

Instead of posting a visualisation task, the client can also interactively modify settings on the back end, such as ordering grid refinements, changing boundary conditions and cell types, or moving geometry through the domain. In this way, an interactive steering of the code is possible while at the same moment an interactive visualisation is able to switch through different scales from very large to very fine.

\subsubsection{Client-Side Visualisation}

The current visualisation is realised using ParaView \cite{paraview}, an open source scientific visualisation tool capable of handling large data sets as well as parallel rendering and much more. ParaView internally uses the visualisation toolkit library VTK \cite{vtk}, an open source library for 3D computer graphics, image processing, and visualisation. 
In order to connect the visualisation immediately to the simulation side, a ParaView plug-in was written, incorporating the actual client implementation.

\begin{figure}[htb]
\centering
	\includegraphics[width=\textwidth]{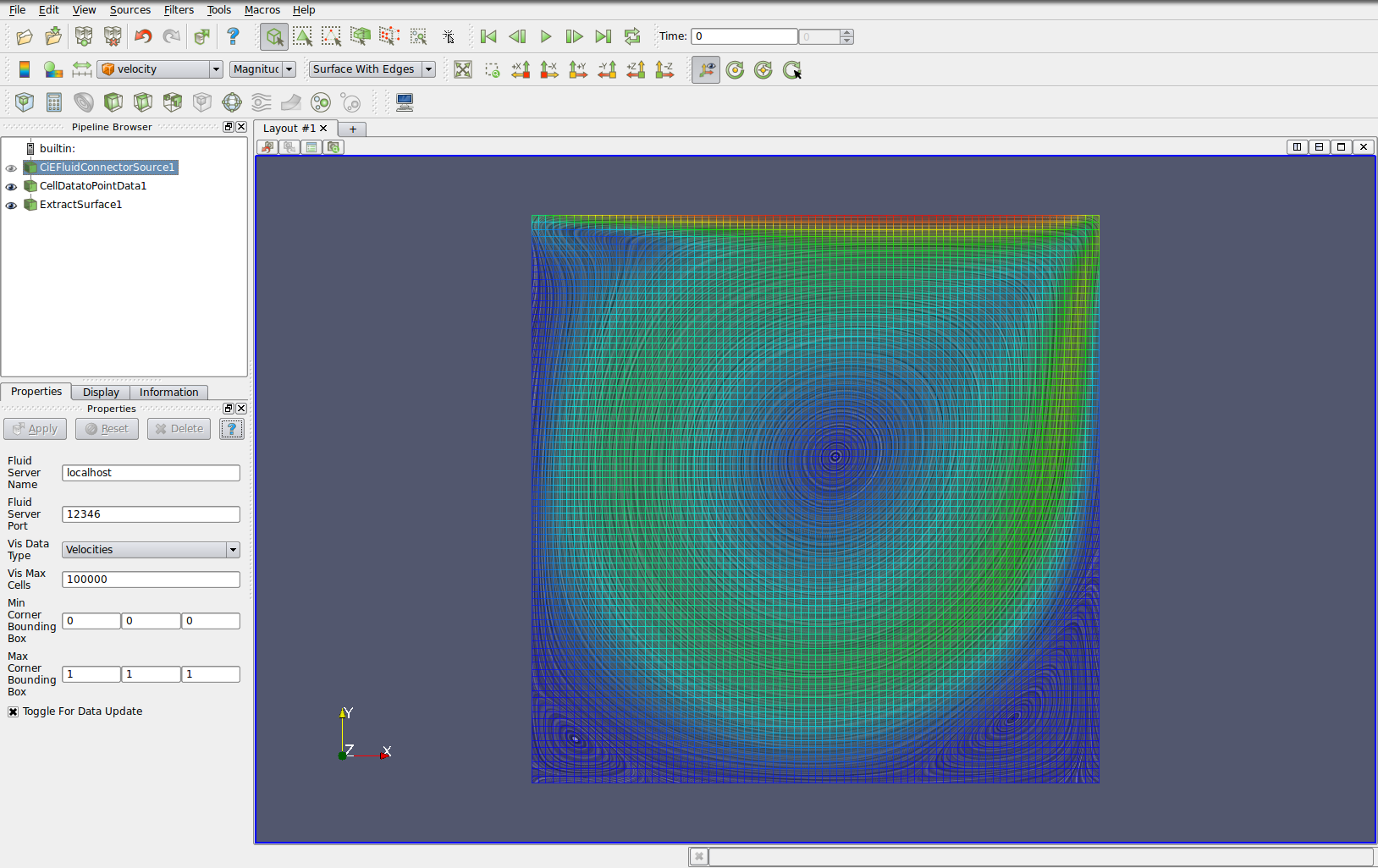}
	\caption{ParaView as visualisation front end enriched by a plug-in for the direct connection to the simulation results computed on the back end (cluster or supercomputer).}
	\label{fig:paraview}
\end{figure}

Figure~\ref{fig:paraview} shows a screen shot of the ParaView GUI with the plug-in details on the lower left side. The user can enter the fluid server name and port as well as the data type to transmit, the maximum amount of cells to be displayed, and the bounding box coordinates of the sliding window visualisation. The plug-in connects via TCP sockets to the server and checks some prerequisites, such as the endianness and size of standard data types. If all prerequisites match, a connection can be established, and among other tasks, a visualisation can be requested (as described in the previous section). The bounding box coordinates, as well as the visualisation data type, and amount of cells to be displayed are sent to the back end, which queries internally for the right data and delivers the results back to the front end for visualisation. Internally, the plug-in receives a long data stream which gets decoded and stored in the internal VTK unstructured grid data structure which then can be immediately visualised by ParaView.

\subsubsection{Performance Measurements}

In order to classify the presented approach, some time measurements were performed. Table \ref{tab:time_perf_meas} gives an overview over update times from the point of request until the data was transferred completely from the server-side collector to the client-side ParaView front end.

\begin{table}[htbp]
  \centering
  \caption{Request time in [s] for different problem sizes depending on the window size.
}
    \begin{tabular}{|r||r|r|r|r|}
    \hline
     & \multicolumn{4}{c|}{problem size} \\
     \hline
    window size  & {256$\times$256} & {512$\times$512} & {1024$\times$1024} & {2048$\times$2048} \\
    \hline \hline
    2048$\times$2048  & $-$ & $-$ & $-$ & 45.432 \\
    1024$\times$1024  & $-$ & $-$ & 13.355 & 12.287 \\
    512$\times$512  & $-$ & 3.652 & 3.554 & 6.507 \\
    256$\times$256  & 1.691 & 1.656 & 3.011 & 3.106 \\
    128$\times$128  & 1.169 & 1.168 & 1.164 & 3.001 \\
    64$\times$64  & 1.048 & 1.061 & 1.047 & 3.229 \\
    \hline
    \end{tabular}%
  \label{tab:time_perf_meas}%
\end{table}%

Thus, for different problem sizes the times were measured for visualising different parts of the domain. Hence, in case of a domain with size 1,024$\times$1,024 cells, 13.4 seconds were necessary in order to request velocity data (3 double-precision values plus geometry and topology information) for visualising the complete data set. If only a fourth of the complete data set is of interest (i.\,e. 512$\times$512), the request could be completed in under 3.6 seconds.

Unfortunately, a huge levelling-off can be observed for a low number of cells. This can be explained easily, when regarding the transmission and request process in more details. As stated in Section \ref{ssec:sliding_window_implementation}, the client submits a request to the server-sided collector, which forwards the necessary information to the neighbourhood server. In the current implementation, the clients check for changes occurring due to the neighbourhood server only once at the beginning of every time step, thus operations within a certain time step are currently not interrupted by visualisation requests. Hence, the measured update time cannot be smaller than the actual time necessary for one step to complete. These times, however, are again strongly dependent on the number of processes used for the simulation.


\begin{figure}[!htb]
	\centering
  \includegraphics[height=14cm, angle=-90]{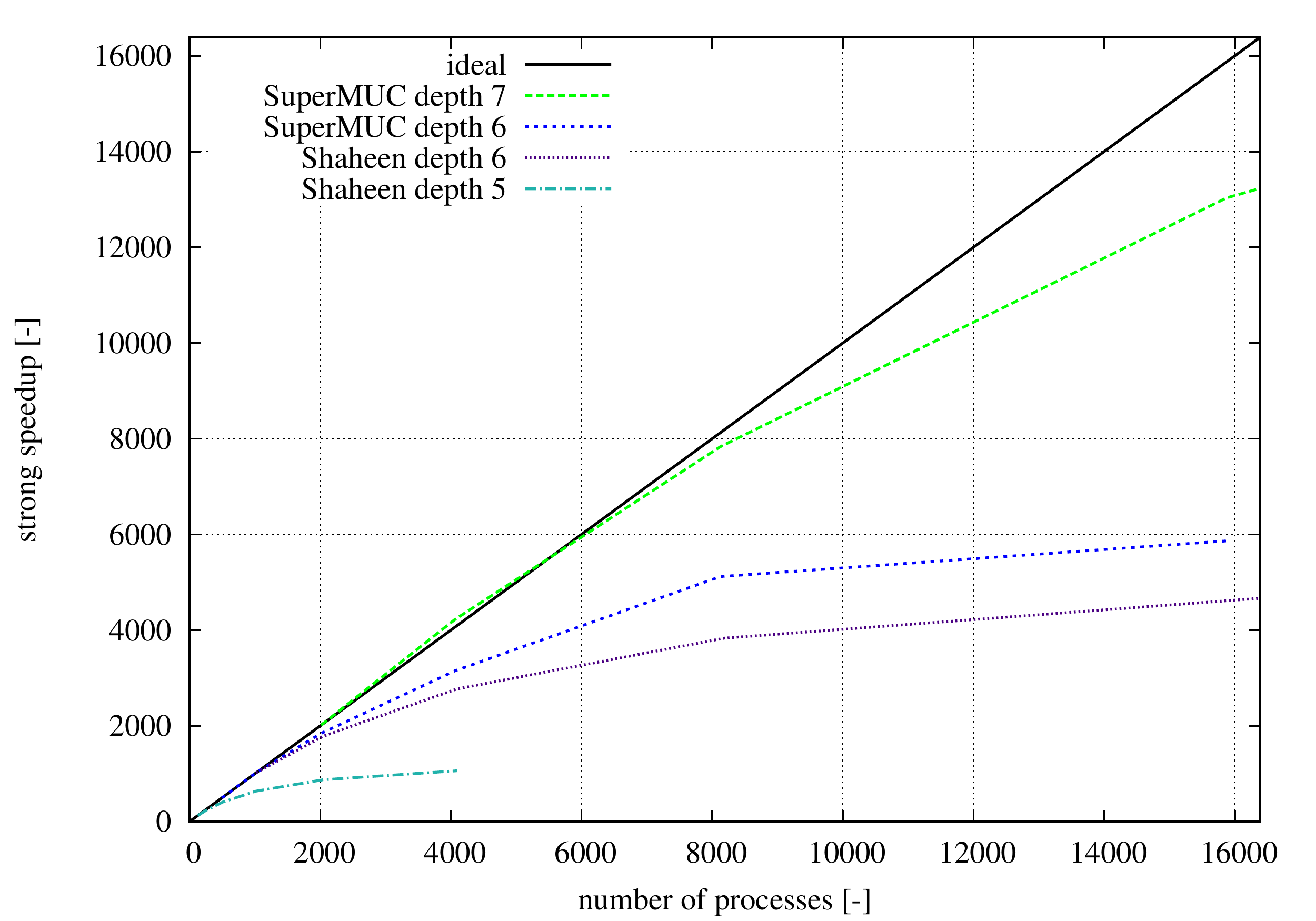} 
	\caption{Strong speedup results on two supercomputing platforms for a 3D domain of fully refined grids with a refinement level of (2,2,2) up to depth 5, 6, or 7 and a discretisation of (16,16,16) cells per grid.}
	\label{fig:impl:speedup}
\end{figure}

For first benchmarks, several small and large size clusters were available. On the small end, a four node (Intel Sandy Bridge) cluster with 64 cores installed at the Chair for Computation in Engineering in Munich as well as the MAC-Cluster\footnote{MAC-Cluster: \texttt{http://www.mac.tum.de/wiki/index.php/MAC\_Cluster}} installed in the Munich Centre of Advanced Computing were used -- the latter one for extensive testing of the sliding window concept using up to 512 processes. The simulation code itself was tested (without the sliding window collector server) on SuperMUC\footnote{SuperMUC: \texttt{http://www.lrz.de/services/compute/supermuc}}, one of Germany's national supercomputers using more than 140,000 processes for simulation purposes, as well as on Shaheen\footnote{Shaheen: \texttt{http://ksl.kaust.edu.sa/Pages/Shaheen.aspx}}, a 16 rack BlueGene/P installed at King Abdullah University of Science and Technology (KAUST) in the Kingdom of Saudi Arabia. Figure \ref{fig:impl:speedup} shows speedup measurements of the pure simulation code for a 3D example similar to the lid driven cavity shown in Figure \ref{fig:re3200:sliding_window_cell_selection} up to 16,000 processes. Depth 5 has 153.4 million cells, depth 6 has 1.23 billion cells, and depth 7 has 9.8 billion cells in total. 

\section{CFD Results}
\label{sec:results}

In the following, we will present two different applications of the sliding window technique for the interactive data exploration of CFD results and for the steering of large-scale simulations as to be observed in flooding scenarios of urban regions.

\subsection{Data Selection and Exploration}

For demonstrating the capabilities of the sliding window concept, the standard lid driven cavity benchmark was chosen. It is described in \cite{Ghia1982387} and consists of a domain of one by one meter with four no-slip walls where the top no-slip wall is moving with a constant velocity of one, thus, exciting the internal fluid by shear stress only.

For the moment, turbulent effects are not modelled, and a complete laminar flow behaviour is assumed. For the lid driven cavity example, this is not a problem at all. For larger, more realistic examples, turbulence modelling is inevitable, but will not be dealt within the context of this paper.

As geometrical discretisation, the grid subdivision was chosen to $n_x=n_y=2$ up to level 3 and $nc_x=nc_y=10$ as cell amount.

\begin{figure}[htb]
\centering
\subfigure[complete][Display of max.\ 400 cells in the sliding window bounding box (0;0) to (1;1)]{%
\label{fig:re3200:complete_l1}%
\includegraphics[width=6.5cm]{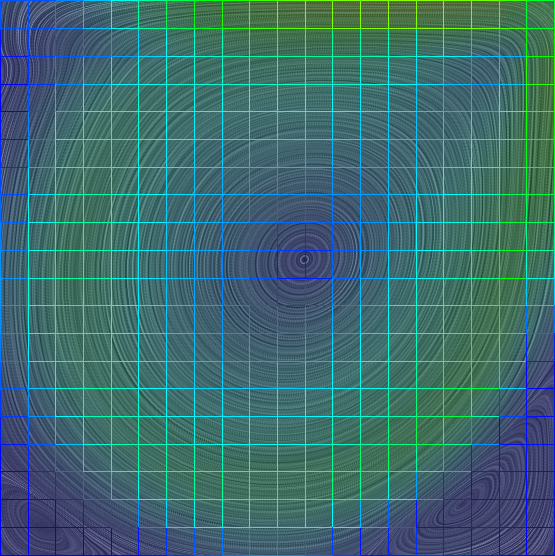} } \quad
\subfigure[half][Display of max.\ 400 cells in the sliding window bounding box (0;0.5) to (0.5;1)]{%
\label{fig:re3200:half_l1}%
\includegraphics[width=6.5cm]{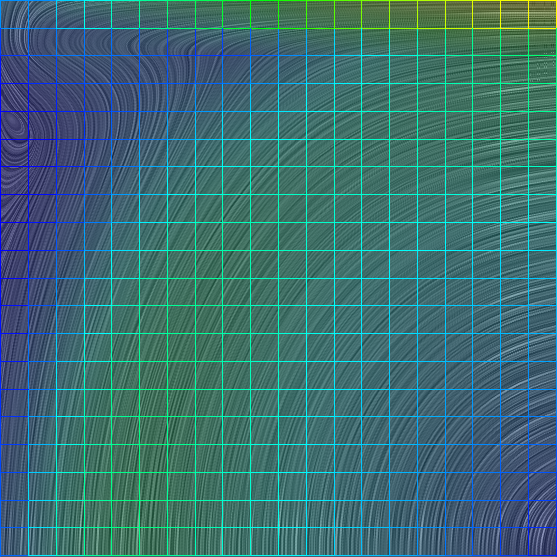} } \\
\vspace{5pt}
\subfigure[quarter][Display of max.\ 400 cells in the sliding window bounding box (0;0.75) to (0.25;1)]{%
\label{fig:re3200:quarter_l1}%
\includegraphics[width=6.5cm]{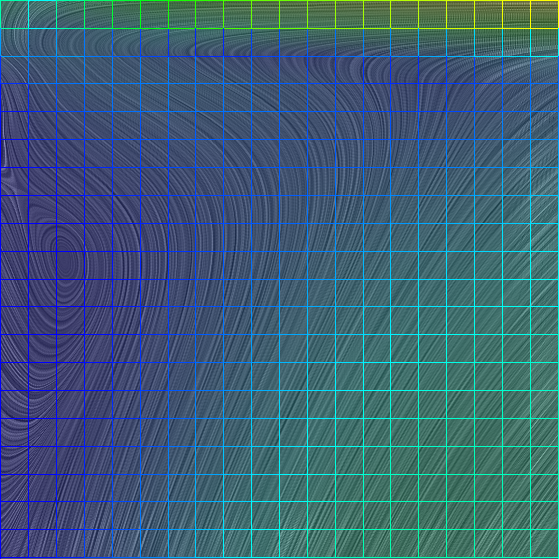} } \quad
\subfigure[legend][Different positions of the selections from a, b, and c]{%
\label{fig:re3200:fig_selection}%
\includegraphics[width=6.5cm]{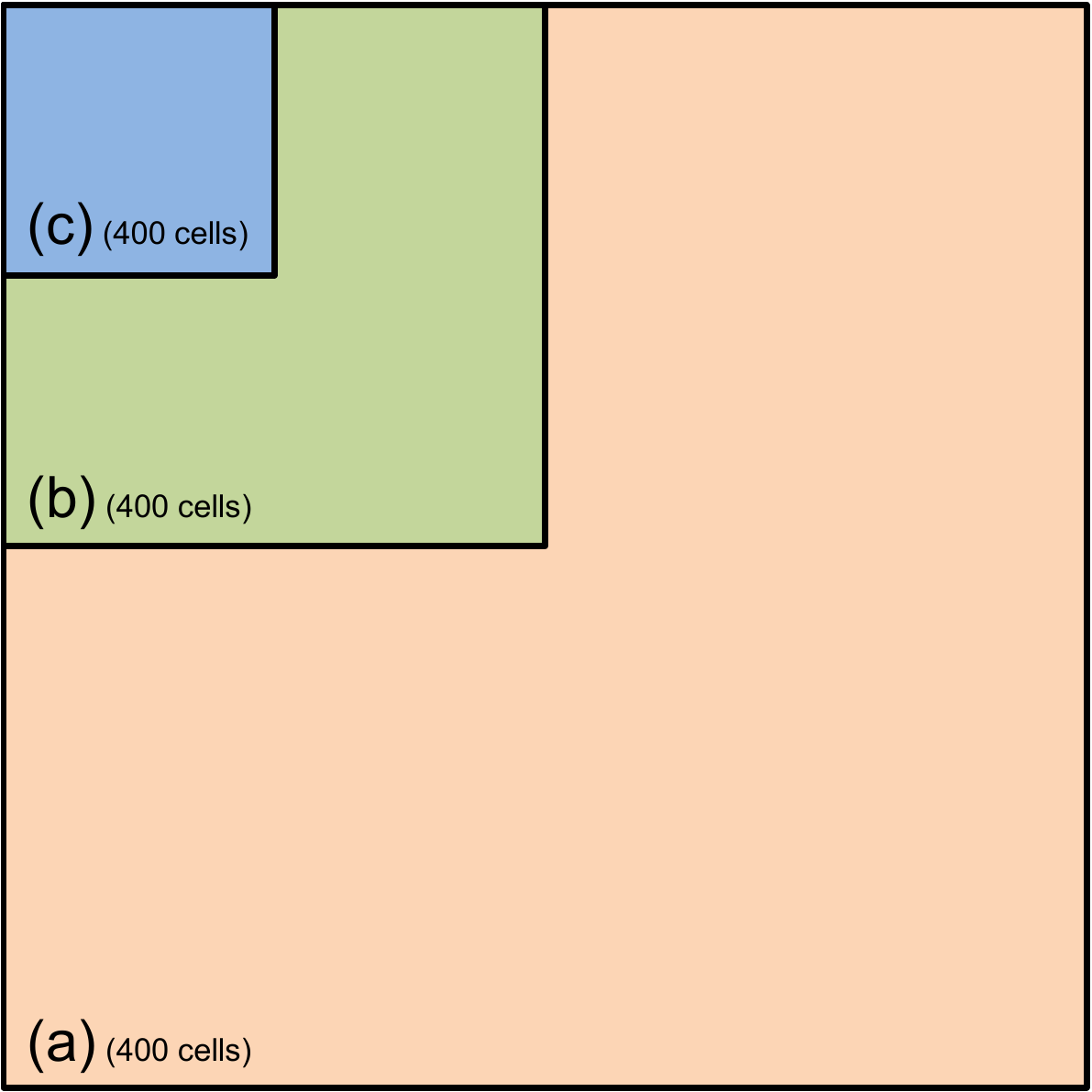} }
\caption{Flow in a lid driven cavity for $Re = 3200$ (visualised using LIC -- Line Integral Convolution), computed with the parallel implementation of our flow solver and visualised by the sliding window concept for different selections, always using a maximum of 400 cells (but always computed on the finest resolution of the computational domain). The vortex in the left upper corner cannot be seen in the coarsest visualisation, only after zooming in.}
\label{fig:re3200:sliding_window_cell_selection}
\end{figure}

Figure~\ref{fig:re3200:sliding_window_cell_selection} shows different visualisation areas selected in the sliding window display keeping the amount of data transferred over the network constant. Figure~\ref{fig:re3200:fig_selection} shows a graphical representation of which bounding boxes, i.\,e.\ windows, were used for data selection. \ref{fig:re3200:complete_l1} shows the complete domain with a coarse resolution of 400 cells. It is very important to stress the fact, that the resolution of the computation is different from the actual resolution of the visualisation given by the sliding window, because the computation is always carried out on the finest grids. After selecting a sliding window by defining a bounding box with half the size in each direction (Figure~\ref{fig:re3200:half_l1}), 400 cells are transferred for visualisation. Hence, the visualisation resolution in the physical domain got higher while reducing the size of the window for display.

In Figure~\ref{fig:re3200:quarter_l1} the size of the sliding window was even reduced further to 25\% of the original size in each direction thus giving even more insight by having a higher physical resolution in this area.

\begin{figure}[htb]
\centering
\subfigure[complete 100][Complete grid with max.\ 100 cells selected]{%
\label{fig:details:complete_l0}%
\includegraphics[width=6.5cm]{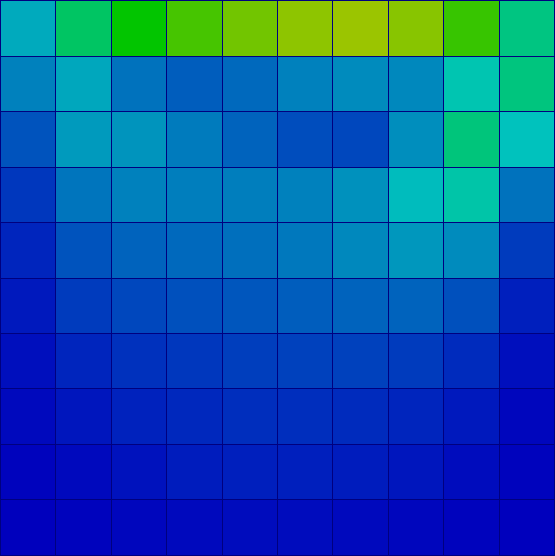} } \quad
\subfigure[complete 400][Complete grid with max.\ 400 cells selected]{%
\label{fig:details:complete_l1}%
\includegraphics[width=6.5cm]{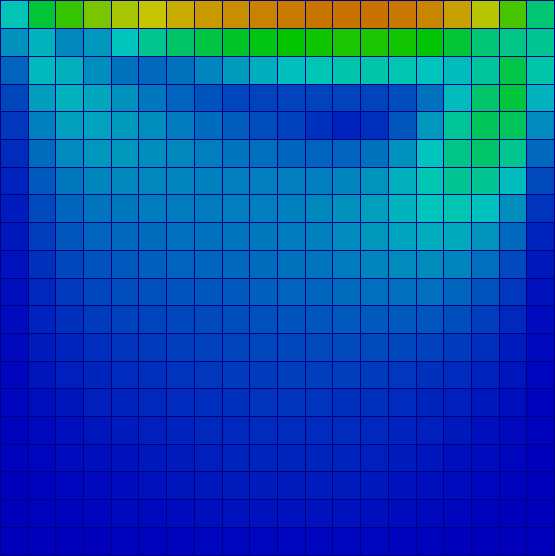} } \\
\vspace{5pt}
\subfigure[complete 1600][Complete grid with max.\ 1600 cells selected]{%
\label{fig:details:complete_l2}%
\includegraphics[width=6.5cm]{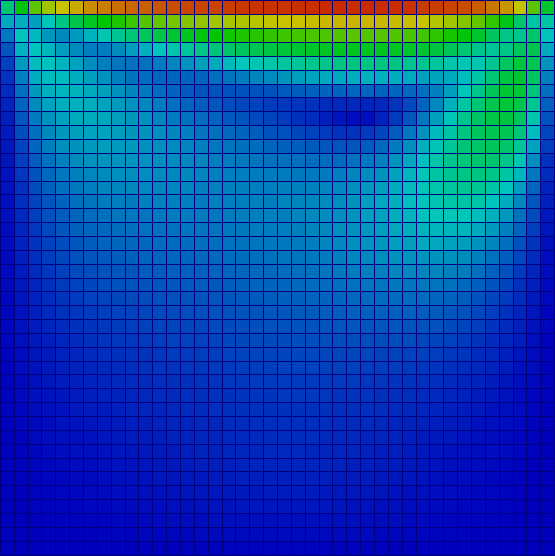} } \quad
\subfigure[complete 6400][Complete grid with max.\ 6400 cells selected]{%
\label{fig:details:complete_l3}%
\includegraphics[width=6.5cm]{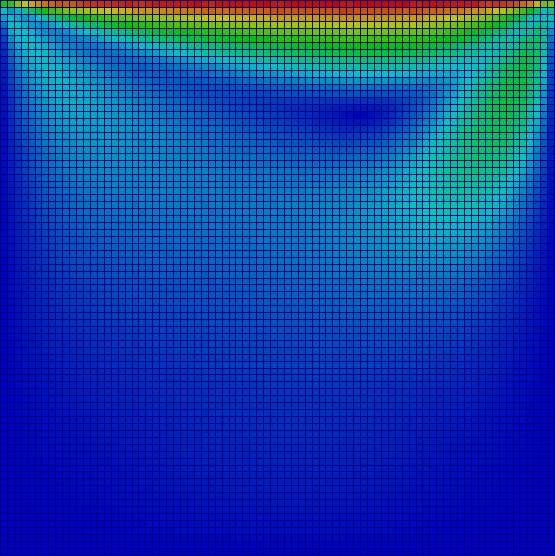} }
\caption{Flow in a lid driven cavity for $Re = 100$, computed with the parallel implementation of our flow solver and visualised for different levels of detail (but always computed on the finest resolution of the computational domain).}
\label{fig:details:sliding_window_cell_selection}
\end{figure}

An additional benefit of this approach -- which arose nearly for free out of the implementation of the sliding window -- is shown in Figure~\ref{fig:details:sliding_window_cell_selection}. Here the sizes of the sliding window itself is not changed, but a different threshold for the cell selection is chosen. This means, that if a maximum of 100 cells should be displayed due to bandwidth limitations, the results look like Figure~\ref{fig:details:complete_l0}. By increasing the threshold continuously, more and more details get visible, until the computational resolution is reached, which is shown in Figure~\ref{fig:details:complete_l3}.

\begin{figure}[htb]
\centering
	\includegraphics[width=14cm]{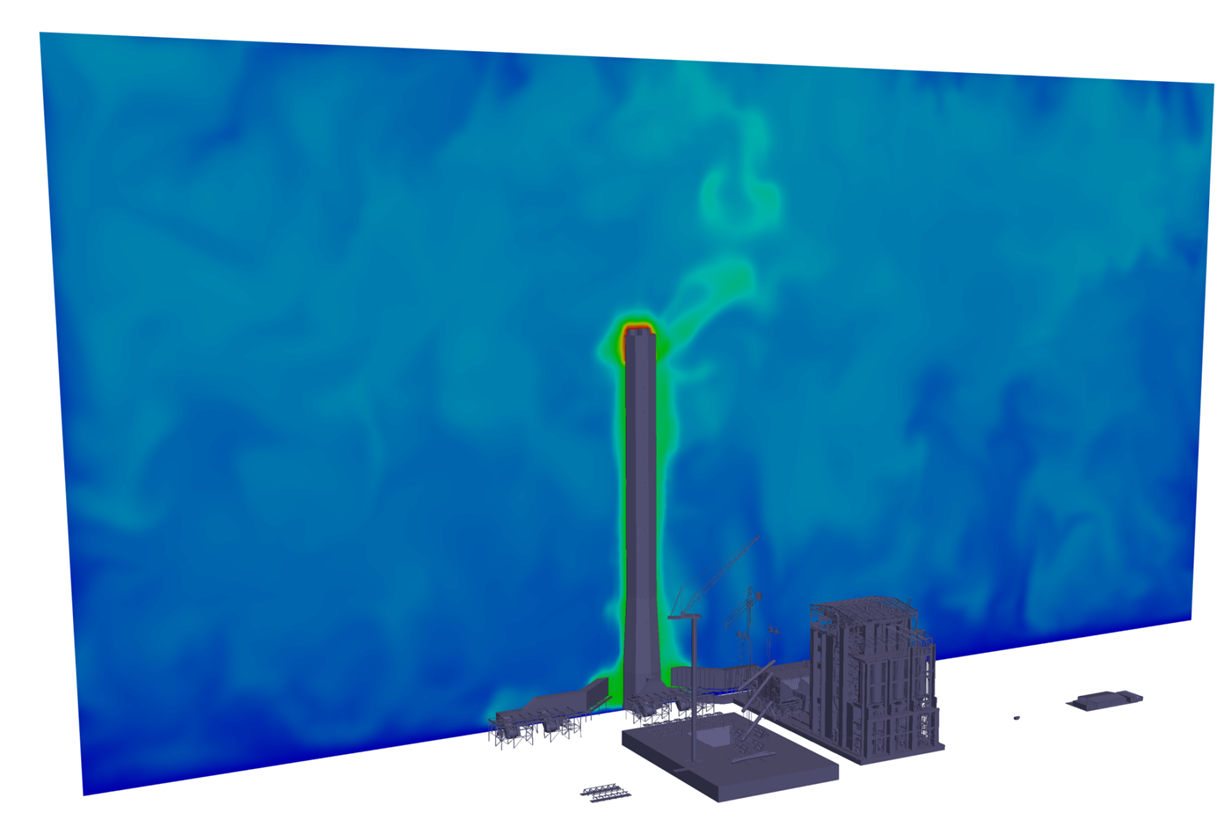}
	\caption{Thermal simulation of a power plant on a very large scale.}
	\label{fig:powerplant_full_scale}
\end{figure}

This easy benchmark simulation now highlights the way and the possibilities the sliding window can be applied to: assume a very large domain containing the power plant depicted in Figure~\ref{fig:powerplant_full_scale}. If an adaptive high performance multi-scale computation should be carried out on hundreds of thousands of processors, it will not be possible to transfer all the data in a reasonable amount of time for visualisation purposes to a dedicated front end node. If the visualisation should be performed in a reasonable amount of time or even in real time, only some data points of the grids will be sent to the visualisation node, thus showing a coarse picture of the complete domain. Assuming the time for sending these visualisation results is sufficient, one then can gradually zoom into the domain while transferring more and more accurate data and omitting less and less points until such a small part of the domain is reached, that every data point can be transmitted and the visual resolution meets exactly the computational one.

\begin{figure}[htb]
\centering
	\includegraphics[width=14cm]{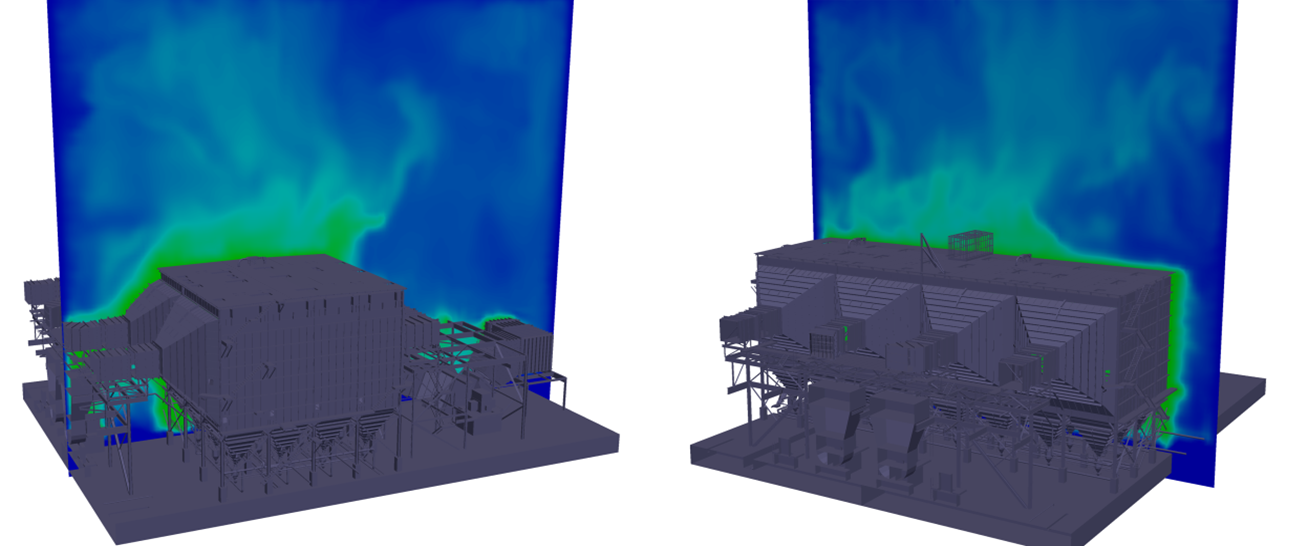}
	\caption{Thermal simulation of a small part of the power plant on a small scale using as much degrees of freedoms as the full scale model.}
	\label{fig:powerplant_small_scale}
\end{figure}

This is presented in Figure~\ref{fig:powerplant_small_scale}. Here, the same visual resolution for displaying a (geometric) detail is used as in the case where the complete power plant model was shown.

\subsection {Steering Large-Scale Simulations}

Another aspect of the sliding window concept concerns the steering of large-scale simulations. Due to the hierarchical approach of our data structure we can -- practically -- generate any desired resolution needed for a further processing. This approach is also known as level-of-detail \cite{Samet:90} and usually comes into play when dealing with multi-scale data. In our case, we have an agglomeration of different data stored into a central data base. This data base comprises GIS, i.\,e.\ geoinformation, and BIM, i.\,e. Building Information Modeling \cite{Eastman:08} data, the latter one providing \emph{"a digital representation of physical and functional characteristics of a facility \cite{BIM}."}
Here, users have the possibility to navigate through terrain and built infrastructure information for visual exploration in real time \cite{Varduhn:11}, allowing a \emph{nearly} seamless transition from coarse scales (whole cities) to single buildings (including furniture etc.). Based on this mashup, numerical simulation shall now be coupled to multi-resolution geometry using the sliding window concept in order to select regions of interest for multi-scale urban and
environmental simulations.

For instance, any question regarding hydromechanical aspects in urban management or heating, ventilation, and air conditioning (HVAC) of indoor thermal comfort can be answered on the same model. Furthermore, aspects over different scales are also resolvable, allowing for instance to study global and local effects of flooding scenarios. While on the global scale the focus is more on the propagation of water in urban districts, on the local scale questions of water penetration into buildings and estimation of damage are of interest. Applying the sliding window technique to such scenarios offers a possibility of steering the underlying simulation, namely selecting subsets of the scene during application runtime for a new computation on a finer resolution. In contrast to the power plant simulation from above (see Figures~\ref{fig:powerplant_full_scale} and \ref{fig:powerplant_small_scale}) where a computation on a fine resolution was carried out for the entire domain, here only parts of the domain are re-computed on a finer scale.
As those fine-scale computations can be very complex, the time between updates might grow large and the effect of real time becomes not evident. Nevertheless, the sliding window concept still allows users to select regions of interest during an online simulation (on the coarse scale) for an offline simulation, i.\,e.\ batch processing, on the local scales for exploration.

The idea is, to start with a simulation on the coarse scale (i.\,e.\ entire urban districts) where the window initially covers the entire domain. Then the user can resize and move the window to any part of the domain he or she is interested in exploring details. Once the simulation results on the coarse scale are available they will be used as boundary conditions for the fine scale computation, thus leading to a unidirectional coupling of the different scales. It is obvious, that a bidirectional coupling becomes necessary to model a correct transportation of physical quantities in both directions.
First numerical experiments with transient boundary conditions were very promising \cite{varduhn:14}, and thus
underline the potential of this approach.

\begin{figure}[htb]
\centering
	\includegraphics[width=\textwidth]{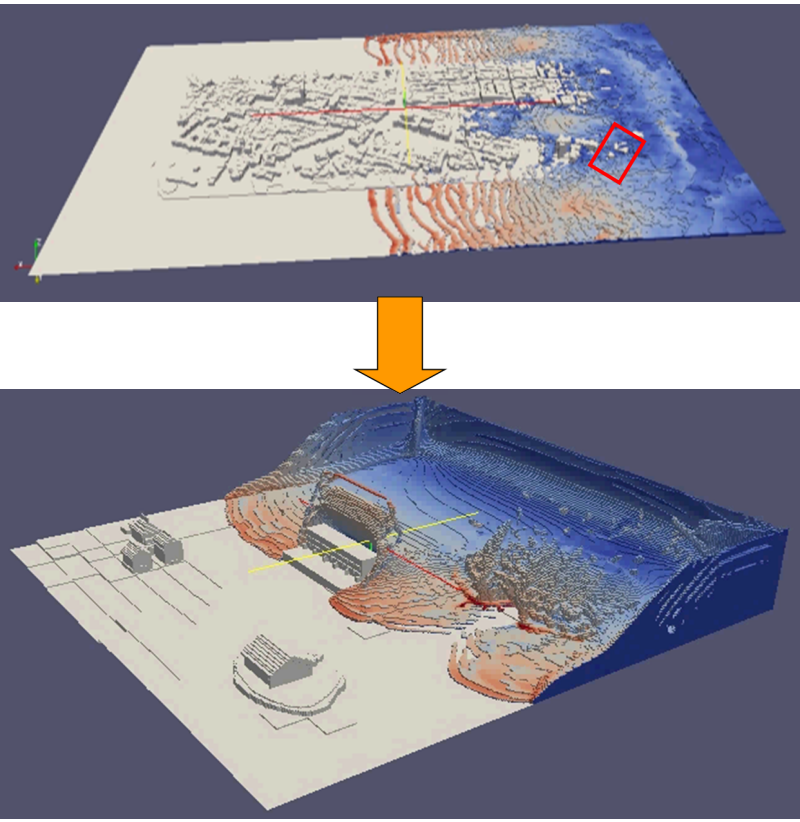}
	\caption{Steering of a multi-scale simulation with sample application of a flooding scenario: here, interactive selection of region in coarse simulation (entire domain) via sliding window (red box in upper picture) for steering the simulation, i.\,e.\ re-computation of smaller region on finer resolution in order to explore local effects (overflow of a single building; lower picture)}
	\label{fig:flooding}
\end{figure}

Figure~\ref{fig:flooding} shows the flooding of an urban district and the overflow of a single building selected by the sliding window---here, the interaction happens online (i.\,e.\ during run time) but the computation runs mostly offline (as batch job) due to the computational complexity of the simulation.

Due to the sliding window concept the resolution on all scales stays nearly constant, anyhow the computational complexity on finer scales increases as time steps between two updates become smaller and smaller. While a coarse scale representation of the entire domain might still be computed in real time on a cluster or small supercomputing system, the fine scales typically entail the usage of more computing resources which -- due to commissioning policies of supercomputing centres -- are unlikely available during simulation runtime. Hence, different strategies are necessary to leverage interactive high-performance computing for the emerging field of high-fidelity simulations.
For instance, some supercomputing centres offer the possibility for interactive, i.\,e.\ online computations. Here, users get exclusive access on computing resources, hence any direct conection between front-end and back-end system can be established in order to steer the application via the sliding window concept. In case direct access to the computing nodes -- due to hidden IP addresses, e.\,g.\ -- is not possible, each node has to dump its data stream to a file instead, which is then processed by the collector (running on the master node) and returned to the user.

\section{Conclusion}
\label{sec:conclusion}

In this paper, we have presented a sliding window concept for interactive high-{perform\-ance} computing that allows a user to select and slide/resize a region of interest within the computational domain for the visual display of simulation results in order to restrict and keep constant the data transfer between a simulation back end and visualisation front end. This becomes necessary as any complex simulation with several hundreds of thousands of data points easily exceeds the bandwidth of modern networks and, thus, hinders an interactive processing of the running simulation as required for computational steering. We have further shown first very promising results of this concept for a fluid flow solver, where a user out of ParaView can control and interact with a parallel CFD code, giving him or her the possibility to study both large and small scale phenomenas. Furthermore, we have presented a second application -- a flooding scenario -- of the sliding window concept for steering a multi-scale simulation where the user can select any region of the domain for a re-computation on a finer resolution in order to explore local effects. Such examples underline the need for sophisticated concepts like this in order to interactively process huge data sets and to leverage interactive high-performance computing.

\bibliographystyle{IEEEtran}
\bibliography{bibtex}

\end{document}